# Growth and applications of two-dimensional single crystals

Zhibin Zhang[1#], Stiven Forti[2#], Wanqing Meng[3#], Sergio Pezzini[4#], Zehua Hu[3#], Camilla Coletti[2,5*], Xinran Wang[3*], and Kaihui Liu[1,6*]

[1] State Key Laboratory for Mesoscopic Physics, Frontiers Science Centre for Nano-optoelectronics, School of Physics, Peking University, Beijing, China

[2] Center for Nanotechnology Innovation @NEST, Istituto Italiano di Tecnologia, Pisa, Italy

[3] National Laboratory of Solid State Microstructures, School of Electronic Science and Engineering and Collaborative Innovation Center of Advanced Microstructures, Nanjing University, Nanjing, China

[4] NEST, Istituto Nanoscienze-CNR and Scuola Normale Superiore, Pisa, Italy

[5] Graphene Labs, Istituto Italiano di Tecnologia, Genova, Italy

[6] Songshan Lake Materials Laboratory, Institute of Physics, Chinese Academy of Sciences, Dongguan, China

# These authors contributed equally to this work

* Correspondence: khliu@pku.edu.cn; xrwang@nju.edu.cn; camilla.coletti@iit.it

**Two-dimensional (2D) materials have received extensive research attentions over the past two decades due to their intriguing physical properties (such as the ultrahigh mobility and strong light-matter interaction at atomic thickness) and a broad range of potential applications (especially in the fields of electronics and optoelectronics). The growth of single-crystal 2D materials is the prerequisite to realize 2D-based high-performance applications. In this review, we aim to provide an in-depth analysis of the state-of-the-art technology for the growth and applications of 2D materials, with particular emphasis on single crystals. We first summarize the major growth strategies for monolayer 2D single crystals. Following that, we discuss the growth of multilayer single crystals, including the control of thickness, stacking sequence, and heterostructure composition. Then we highlight the exploration of 2D single crystals in electronic and optoelectronic devices. Finally, a perspective is given to outline the research opportunities and the remaining challenges in this field.**

## 1. Introduction

Since the first isolation of graphene using the simple 'Scotch tape' method in 2004 [1], 2D materials have attracted extensive attention due to their excellent properties in the limit of atomic thickness. Graphene, the first obtained 2D material, combines numerous unique properties, such as high mechanical strength, thermal and electrical conductivity, and optical transparency [2], making it attractive for various applications, including high-performance electronic and optoelectronic devices, flexible transparent devices and protective coatings [3]. These promising properties and application potentials have triggered the exploration of 2D materials beyond graphene. Up to now, several groups of 2D materials have been isolated, including (but not limited to) transition metal dichalcogenides (TMDs) [4], hexagonal boron nitride (hBN) [5] and MXene [6]. Their novel physical and chemical properties and great application prospects substantially expand the frontier of 2D materials. From the viewpoint of the electronic properties, the 2D material family includes a full range of conductors (graphene), semiconductors (TMDs like $MoS_2$, $WS_2$ and $VS_2$), insulators (hBN) and Mott-insulators ($TaS_2$ and $TaSe_2$), which are appealing building blocks for the construction of all-2D electronic and optoelectronic devices. Importantly, their ultimately thin thickness could in principle solve the long-troubling short-channel effect in the post Moore-era.

To realize these attractive applications based on 2D materials, the growth of large-area 2D single crystals is a prerequisite, which has gone through two major stages, i.e., the growth of monolayer and multilayer single crystals. There are several methods developed to realize the growth of 2D materials, such as molecular beam epitaxy (MBE) [7, 8], atomic layer deposition (ALD) [9-11], physical vapor deposition (PVD) [12-14], chemical vapor transport (CVT) [15-17] and chemical vapor deposition (CVD) [18-20]. Among these methods, the CVD technique has shown its advantage in the growth of large-scale 2D materials. Therefore, in this review, we mainly focus on the CVD method. The past decades have witnessed the development of the growth technique for 2D monolayers, from the single-nucleus method (growing of one nucleus into a single crystal) to the multi-nuclei method (stitching of numerous nuclei with the same orientation into a single crystal). At present, the multi-nuclei method has been widely adopted due to its high efficiency, stable controllability, general applicability and better compatibility with industrial production [21]. To achieve the unidirectional alignment of the crystal domains, different kinds of strategies have been developed according to the distinct lattice structure of different materials. Generally speaking, the epitaxy of centrosymmetric materials (e.g., graphene) can be achieved by regulating the interfacial coupling between the substrate and the material [22]; while the epitaxy of non-centrosymmetric materials (e.g., hBN and TMDs) usually requires the substrate to have the same or lower surface symmetry compared with the material [23], which is usually achieved by engineering step edge structures on the substrate surface. Going beyond the monolayer paradigm offers exciting opportunities to extend the capabilities of 2D materials. The power of the multilayer approach stems from the combination of two basic degrees of freedom: (i) the number of layers, and (ii) the stacking sequence. Since the possible stacking sequences of a van der Waals (vdW) crystal also include the twisted configurations, i.e. rotations over the vertical axis by an arbitrary angle, the resulting parameter space is immense [24]. In addition, stacking different materials in vertical heterostructures can further increase the tunability of 2D multilayers [25].

Nowadays, high-performance electronic and optoelectronic applications based on high-quality 2D single crystals are in the spotlight of the scientific community. Moore's Law requires a continuous increase of the integration density in a single chip, while the size reduction of the conventional silicon transistors almost reaches its physical limit. 2D materials can complement conventional bulk semiconductors (such as Si, GaAs, and HgCdTe) in terms of complementary metal oxide semiconductor (CMOS) compatibility, low-cost manufacturing, immunity to lattice mismatch, and flexibility, and thus could further extend Moore's law. It's well known that graphene is unsuitable for logic circuits due to its zero bandgap and the corresponding small current on/off ratio in its field-effect transistors (FETs) [26]. In contrast, TMDs provide an opportunity for new electronic devices such as FETs, as well as optoelectronic devices such as photodetectors and optoelectronic memories, which take advantage of their tunable bandgaps and semiconducting natures [27].

Here, we comprehensively review the single crystal growth methods for the principal 2D materials and their main applicative potentials (figure 1). The methods developed so far for the growth of 2D monolayer and multilayer single crystals are analysed, including their tunability in terms of thickness, stacking order and heterostructure composition. Based on the mature growth technology of 2D single crystals, various high-performance FETs and optoelectronic devices are also explored. Finally, the challenges presently tackled within the field are discussed and the future directions for the growth and applications of 2D materials are envisaged.

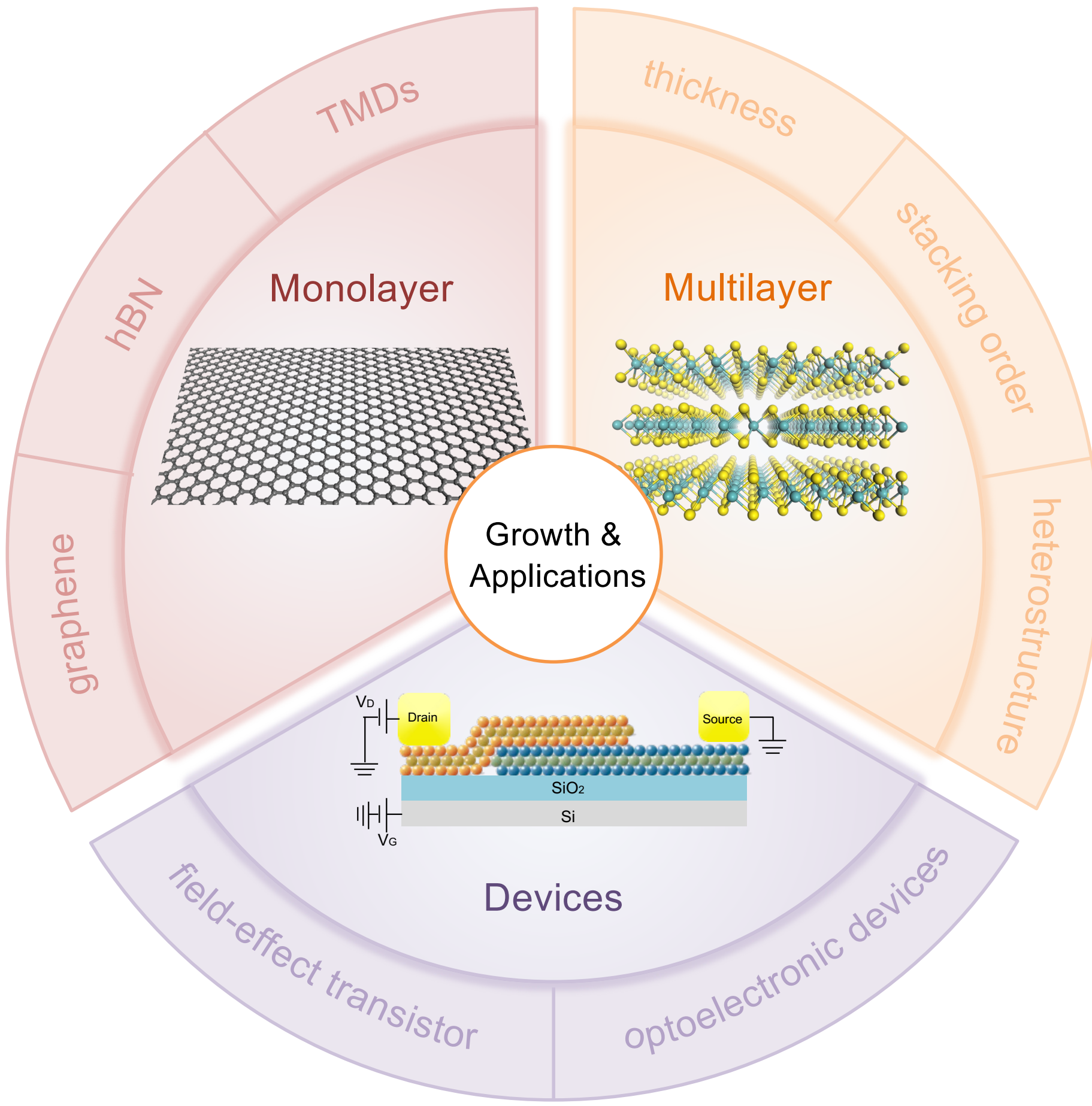

**Figure 1.** Growth and applications of two-dimensional single-crystals.

## 2. Growth of 2D monolayer single crystals

Due to the specific lattice structure of 2D materials (centrosymmetric materials like graphene; non-centrosymmetric materials like hBN and TMDs), their single-crystal growth strategy follows different mechanisms. After decades of efforts, now the epitaxial growth of large 2D single crystals has been successfully demonstrated. In this section, we will discuss these growth strategies sequentially.

### 2.1 Growth strategy for single-crystal graphene

The main approaches for the growth of graphene single crystals are two, i.e., the single-nucleus strategy and the multi-nucleation strategy. The first approach consists in achieving the growth of a large single crystal from a single nucleation centre. The second approach requires one to control the orientation of multiple nucleation centres which then seamlessly stitch into a large single-crystal film.

For the single-nucleus strategy, the control of graphene nucleation density is of vital importance. Generally, there are two main routes that have been developed. One is to reduce or passivate the active sites on metal to suppress the nucleation of graphene. For example, treating Cu with organic solvent [28, 29] (such as acetone and isopropanol) and acid, alkaline or salt solutions [30-34] (such as HCl, HF, $HNO_3$, KOH, or $FeCl_3$, $NH_4S_2O_8$) could effectively remove the surface organic contamination and etch the copper oxide. And then the smoothed Cu surface could be beneficial to achieve a reduced nucleation growth. Through adjusting the annealing parameters (including temperature [35], pressure [36, 37] and gas atmosphere [38-40]), a flattened Cu surface can be also obtained for the growth of graphene films. Besides, one can also passivate the active nucleation sites rather than remove them. Ruoff's group deliberately introduced oxygen to reduce the nucleation sites during the CVD process and they successfully obtained the centimeter-scale graphene domains [30]. With 5 minutes of oxygen exposure, they found that the nucleation density was decreased to ~0.01 $mm^{-2}$, and the individual domain could grow to a diameter larger than 1 cm after 12 hours of growth. They also found that the introduction of oxygen shifted the growth kinetics from edge-attachment–limited to carbon-diffusion-limited. As a result, the graphene domain grown with oxygen turns from typical hexagons to be multibranched or dendritic [30]. In 2016, Xu *et al* developed the oxide-assisted ultrafast graphene growth technique which lowers the energy barrier to the decomposition of the carbon feedstock, and they realized a graphene growth rate of 60 μm $s^{-1}$ on Cu foils [41]. Lin *et al* used melamine pretreatment to control the graphene nucleation on copper surface, and they achieved the growth of centimeter-sized single-crystal graphene. The key point for their successful passivation of the active sites relied on the presence of carbon- and nitrogen-containing compound [42].

The other route is to supply the carbon precursor in a dilute way or a local-feeding way. Li *et al* used a much lower flow rate and partial pressure of methane (less than 1 sccm and 50 mTorr, respectively) to obtain a low density of nuclei, and they finally obtained 0.5 mm large graphene [36]. Gao *et al* changed the flow-rate ratio of $CH_4/H_2$ from 20/400, 10/700 to 5/700 and 4/700 under ambient pressure, and the domain size of the graphene was found to be increased from 50, 100 to 500 and up to 1300 μm [38]. Reference [43] reported that the adoption of ex-situ passivated copper foils together with the reduction of the carbon species impingement flux during growth could yield graphene single crystals of several millimeters in size with growth rates of up to 17.5 μm/min. Based on its advantage of effective suppression of nucleation density, the control of low $CH_4$ flow rate till now has become a commonly adopted strategy for the growth of high-quality graphene. Even in the multi-nucleation strategy, the flow rate of $CH_4$ is usually controlled to a small level.

In 2016, Wu *et al* developed a local feeding method to supply carbon precursors to a desired position of their optimized Cu–Ni alloy substrate [20]. They first prepared $Cu_{85}Ni_{15}$ alloys by electroplating Ni film on Cu substrate followed by thermal annealing. Then they used a quartz nozzle to locally feed the carbon precursor only at a tiny area (figure 2(a)), and finally achieved a ∼1.5-inch, hexagonally shaped single-crystal graphene in 2.5 h at 1,100 °C (figure 2(b)). In 2018, Vlassiouk *et al* demonstrated a local feeding approach with a moving substrate geometry [44]. They introduced a high-velocity buffer gas to ensure the creation of a sharp concentration gradient at the front and the elimination of undesired seed formation. Then they smoothly pulled the substrate at a sufficiently slow rate (maximum rate of 2.5 cm $h^{-1}$) and obtained the 1-foot-long single-crystal graphene film (figure 2(c)). Although the single-nucleus strategy is effective for large single domain growth, it suffers from low efficiency due to the typical slow growth rate. Even with the help of external pulling force, the productivity still could not meet the high requirements for large-scale industrial applications. Thus, to realize fast growth of large single crystals, one needs to consider the multi-nucleation strategy.

The key for the multi-nucleation strategy lies in the simultaneous formation of numerous nuclei and then seamlessly stitching into a single-crystal film. In 2012, Geng *et al* employed a liquid Cu surface to eliminate the grain boundaries in solid polycrystalline Cu, which resulted in a uniform nucleation distribution and enabled self-assembly of graphene crystals on the liquid surface (figure 2(d)) [45]. Figure 2(e) shows the obtained perfectly ordered 2D lattice structures of graphene domains with similar size on liquid Cu. Besides the liquid metal surface approach ensuring a smooth stitching [46, 47], another important route for the growth of large graphene single crystals is the production of single-crystal metallic substrates with a lattice symmetry similar to that of graphene [32, 46, 48-55].

The Cu(111) facet shares a small mismatch in lattice structure with graphene, and thus is considered to be the most promising substrate for the growth of single-crystal graphene films. Due to the urgency for the growth of large single-crystal graphene films, the preparation methods for single-crystal Cu have developed a lot recently [22, 56, 57], with its size going from millimetre-scale to metre-scale, and more than 35 facets have been reported. In 2017, Xu *et al* developed a temperature gradient driven method for the preparation of Cu(111) [22]. They tapered one edge of the Cu foil to ensure the nucleation of only one Cu(111) grain, then they slid the foil through the hot zone to drive the movement of the grain boundaries till the entire Cu foil transformed into a single crystal (figure 2(f)). With this large single-crystal Cu(111) foil, they employed their oxide-assisted ultrafast graphene growth technique [41] and the principle of the roll-to-roll method [58, 59] to realize the fast growth of single-crystal graphene films. Finally, they obtained a (5 × 50) $cm^2$ monolayer graphene film in only 20 min. The reported alignment level of the graphene domains is larger than 99% (figure 2(g)).

Another multi-nucleation strategy consists in defining the nucleation centers of single-crystal graphene by patterning a thin disk of chromium on the Cu polycrystalline foil. In this way, Miseikis and coworkers have demonstrated the possibility to realize large arrays of single-crystal monolayer graphene [60]. The technique is easily scalable and showed great promise for industrial applications [61]. Compared with the single-nucleus strategy, the multi-nucleation strategy demonstrates high efficiency, stable controllability, general applicability and better compatibility with industrial production. By selecting a suitable substrate, single-crystal 2D materials can be achieved through the seamless stitching of these nuclei. Therefore, the multi-nucleation strategy is now much more adopted and is appealing for the scalable production of 2D material.

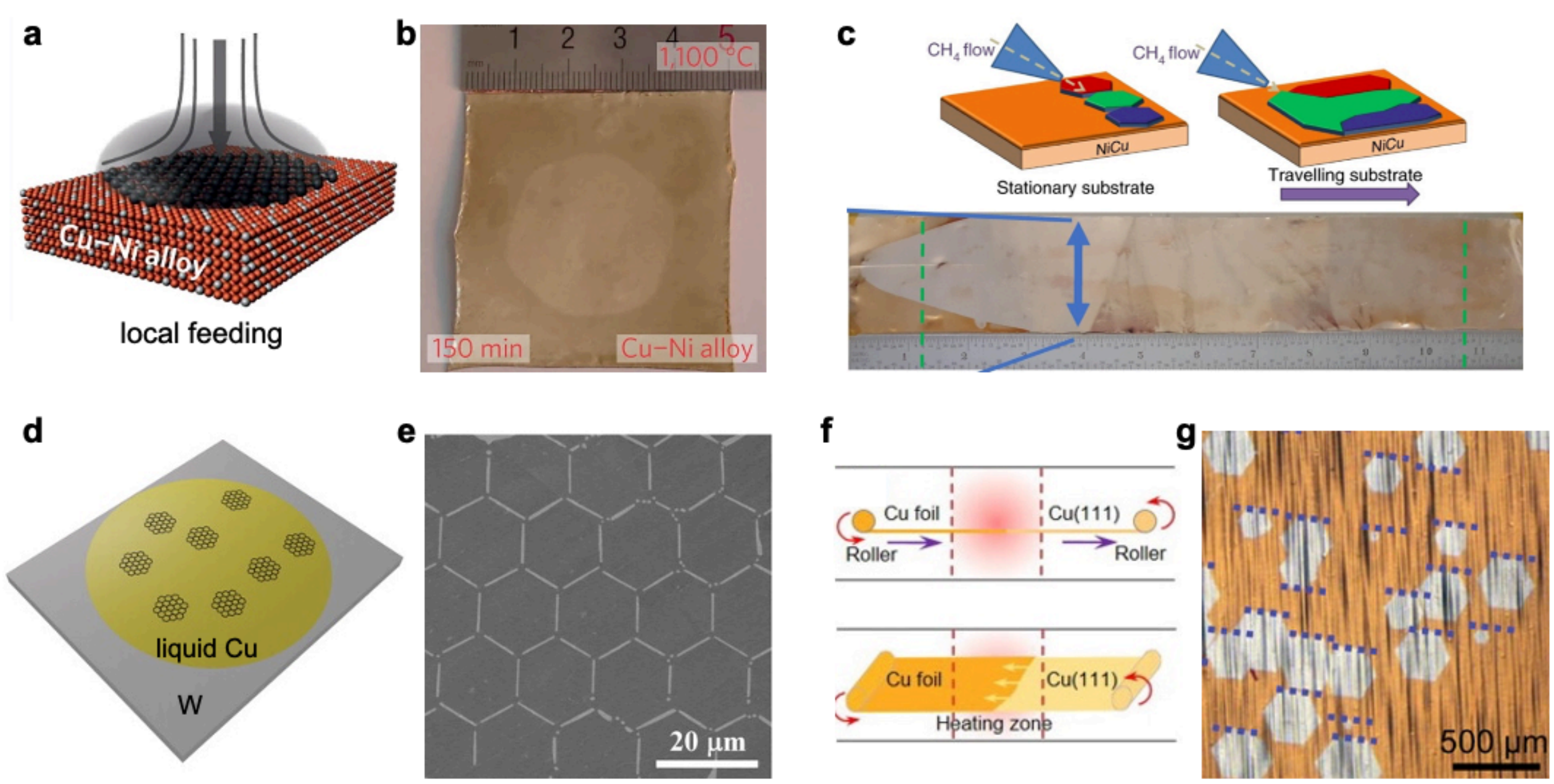

**Figure 2.** Single-nucleus and multi-nucleation strategy for the growth of single-crystal graphene. (a) Schematic for the local feeding strategy [20]. (b) An ∼1.5-inch graphene film grown at 1,100 °C in 150 min on $Cu_{85}Ni_{15}$. Reproduced from [20]. (c) Schematic for the local feeding strategy with travelling substrate and the photograph of the 1-foot-long single crystal graphene. Reproduced from [44]. (d) Schematic for the growth of graphene domains on liquid Cu surface [45]. (e) SEM image of a near-perfect 2D graphene lattice. Reproduced from [45]. (f) Schematics for the continuous production of single-crystal Cu(111) foil [22]. (g) Optical images of aligned large graphene islands. Reproduced from [22].

### 2.2 Growth strategy for single-crystal hBN

The thick BN single crystals are conventionally obtained through high-pressure [62-64], atmospheric-pressure [65], metal flux [66, 67] and vapor-liquid-solid [68] method. To obtain the monolayer single crystals, the CVD method is much more suitable. However, different from graphene, hBN consists of two distinct atoms (boron and nitrogen) in one primitive cell, and thus is a non-centrosymmetric material. Therefore, when hBN is grown on a centrosymmetric surface, the two antiparallel hBN domains have a degenerate formation energy, always resulting in the formation of twin boundaries. To overcome this difficulty, several methods have been proposed.

In 2018, Lee *et al* reported the growth of wafer-scale single-crystal hBN via self-collimated grain formation [19]. They used liquid Au as the growth substrate, and the key is to retain a flat liquid Au with high surface tension to allow for the strong adhesion of borazine precursors. The low solubility of B and N in liquid Au ensures the formation of circular hBN grains. When two hBN grains merge, they will rotate with respect to each other due to the attractive Coulomb interaction between B and N atoms (figure 3(a) and (b)), and these grains will finally seamlessly stitch into a single-crystal film on a wafer scale. Besides, they also demonstrated the direct growth of vertical single-crystal graphene/hBN heterostructure and single-crystal $WS_2$ film.

In 2019, Wang *et al* proposed a more general "edge-coupling-guided" growth strategy for the growth of single-crystal hBN [23]. In their work, they prepared a $10 \times 10$ cm$^2$ single-crystal Cu(110) vicinal surface as the growth substrate which is tilted slightly from the ideal one, with a tilt degree less than 1˚. The surface of Cu(110) consists of parallel step edges along the <211> direction, which reduce the surface symmetry of Cu substrate to a $C_1$ symmetry. This enabled the coupling of Cu<211> step edges with hBN zigzag edges, resulting in the unidirectional alignment of millions of hBN nuclei over a large area (figure 3(c) and (d)). Later, Wu *et al* demonstrated the fabrication of a single-crystal Cu library, and they successfully realized the growth of single-crystal hBN films on different high-index Cu foils [57].

Besides adopting step edges on Cu(110), Chen *et al* also demonstrated the epitaxial growth of two-inch single-crystal hBN films on Cu(111) via lateral docking of hBN to Cu (111) steps [69]. They sputtered a 500-nm-thick polycrystalline Cu film on c-plane sapphire wafers and annealed it at a high temperature (1,040–1,070 °C) with hydrogen to achieve single-crystal Cu(111) films. With the presence of step edges on the Cu(111) surface, ensuring the mono-orientated growth of large hBN films (figure 3(e) and (f)). Furthermore, as the hBN film is grown on a Cu/sapphire wafer, they readily realized the detachment of hBN by polymer-assisted transfer with the help of electrochemical processes, which is promising for the integration of 2D-based transistors.

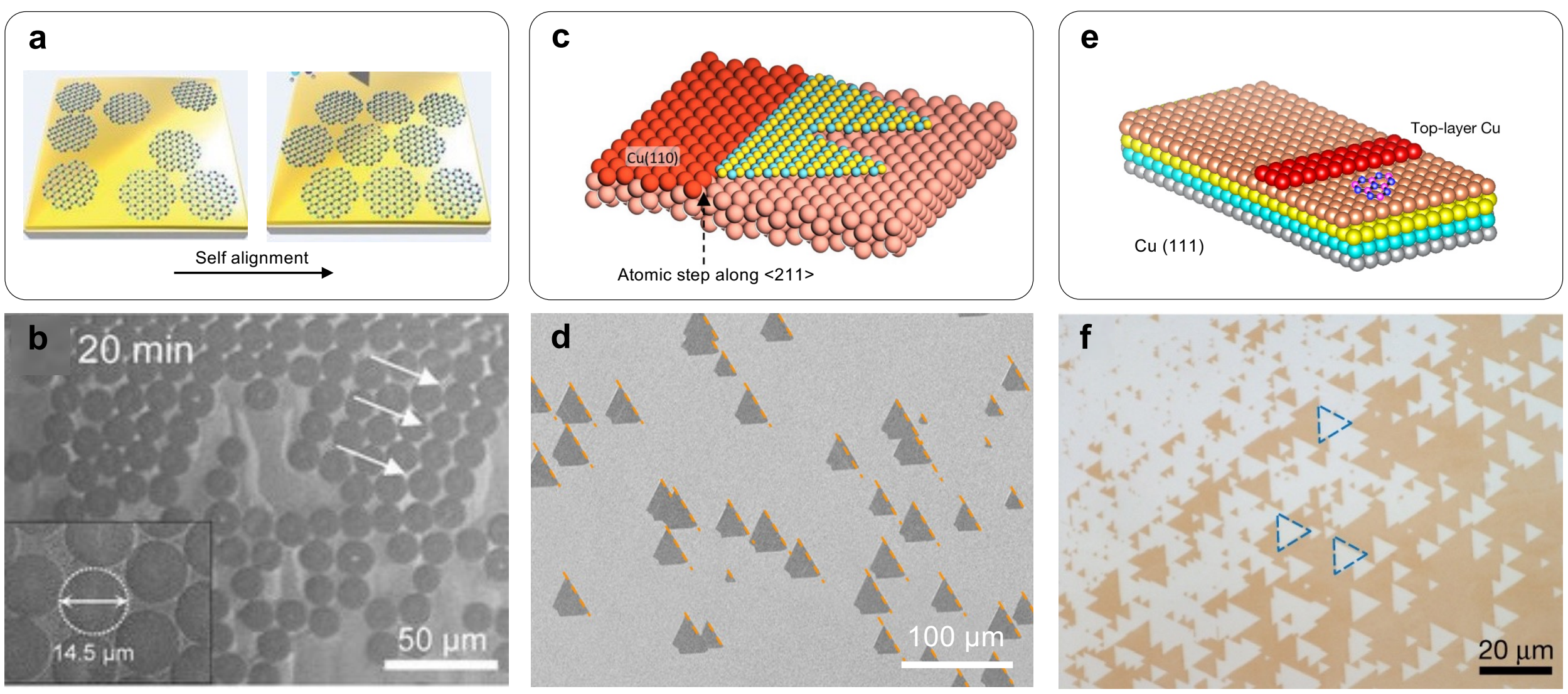

**Figure 3.** Growth strategy for single-crystal hBN. (a) Schematic illustration for the growth of single-crystal hBN film by self-collimated circular hBN grains [19]. (b) Circular hBN grains formed after the growth of 20 min. Reproduced from [19]. (c) Schematic for the Cu(110) surface [23]. (d) SEM image of the as-grown unidirectionally aligned hBN domains on the Cu (110) substrate. Reproduced from [23]. (e) Schematic for the of the BN–Cu(111) configuration [69]. (f) Mono-oriented hBN domains on single-crystal Cu(111) films. Reproduced from [69].

### 2.3 Growth strategy for single-crystal TMDs

In the past decades, approaches for the growth of large continuous TMD films have seen a significant development, yet the obtained TMDs were, until recently, mostly polycrystalline [18, 70-75]. To address this issues, great efforts have been made and several progresses have been reported in very recent times.

In principle, TMDs have similar lattice structure as hBN, thus one would suppose that single-crystal TMDs could also been grown by the design of step edges on substrate. However, its realization is not such straightforward. The most common commercially available sapphire substrate is the C-sapphire wafer with a miscut angle towards the M axis (also C/M) [76], whose steps are along the <11$\bar{2}$0> direction. As shown in figure 4(a), the <11$\bar{2}$0> direction is parallel to the armchair (AC) edge of the triangular TMD domains. The formation energies for two antiparallel TMD domains along the AC edge are calculated to be very close (figure 4(b)), which could lead the formation of twin boundaries. To address this issue, they custom-designed C-sapphire wafers with a major miscut angle towards the A axis (also C/A) for $MoS_2$ growth (the cutting angles ranging from 0.25 to 0.89°), with the steps along the <10$\bar{1}$0> direction. Theoretical calculations revealed that the ZZ-Mo edge with a 100% S coverage (ZZ-Mo-$S_2$) is the most stable configuration under the S-rich condition, with formation energy ~0.1 eV $Å^{-1}$, lower than that of the opposite domain orientation (ZZ-$S_2$, figure 4(b)). Through the change of miscut orientation from C/M to C/A of sapphire, they broke the degeneracy of nucleation energy for the antiparallel $MoS_2$ domains and finally led to a more than 99% unidirectional alignment of the $MoS_2$ domains (figure 4(c)).

At the same time, Wang *et al* found that A-plane sapphire is also applicable for the epitaxial growth of single-crystal $WS_2$ monolayer films [77]. They employed a single-crystal A-plane sapphire substrate with a small cutting angle (~0.1°) from the (11$\bar{2}$0) surface for the growth. Before the growth, they annealed the A-sapphire substrates in oxygen to stabilize the atomic steps. Owing to the coupling between $WS_2$ and A-plane sapphire, two energetically degenerated antiparallel $WS_2$ islands could form on the surface of A-sapphire; with the assistance of the designed sapphire step edges, the degeneracy of the two antiparallel $WS_2$ islands was broken and unidirectionally aligned $WS_2$ domains could be obtained. Their calculations also demonstrated that, with the presence of step edges, the energetically favorable state will reduce to only one direction (figure 4(d) and (e)). To evaluate whether the film is formed by unidirectionally aligned $WS_2$ domains, they etched the as-grown $WS_2$ films using $O_2$ and numerous parallelly aligned holes are shown, demonstrating the high single crystallinity (figure 4(f)). Because there are two guiding principles in the growth mechanism, they named the growth method as a "dual-coupling" strategy and they successfully applied it for the epitaxy of $MoS_2$, $WSe_2$ and $MoSe_2$ single crystals.

Moreover, other 2D materials and heterostructures with exotic atomic and band structure are good platforms to study quantum transport phenomena. For example, $TaS_2$ and $TaSe_2$ have stimulated great research interest into the Mott-insulator correlated charge-density-wave study [78, 79]. Recently, the controllable growth of monolayer $TaS_2$ and $TaSe_2$ have been reported through chemical growth method [80, 81], which facilitates the study of the charge-density-wave state down to the monolayer limit.

In summary, by engineering the surface symmetry of the growth substrate, the single-crystal 2D monolayers have been obtained, which provides opportunities for their wide applications in electronic and optoelectronic devices.

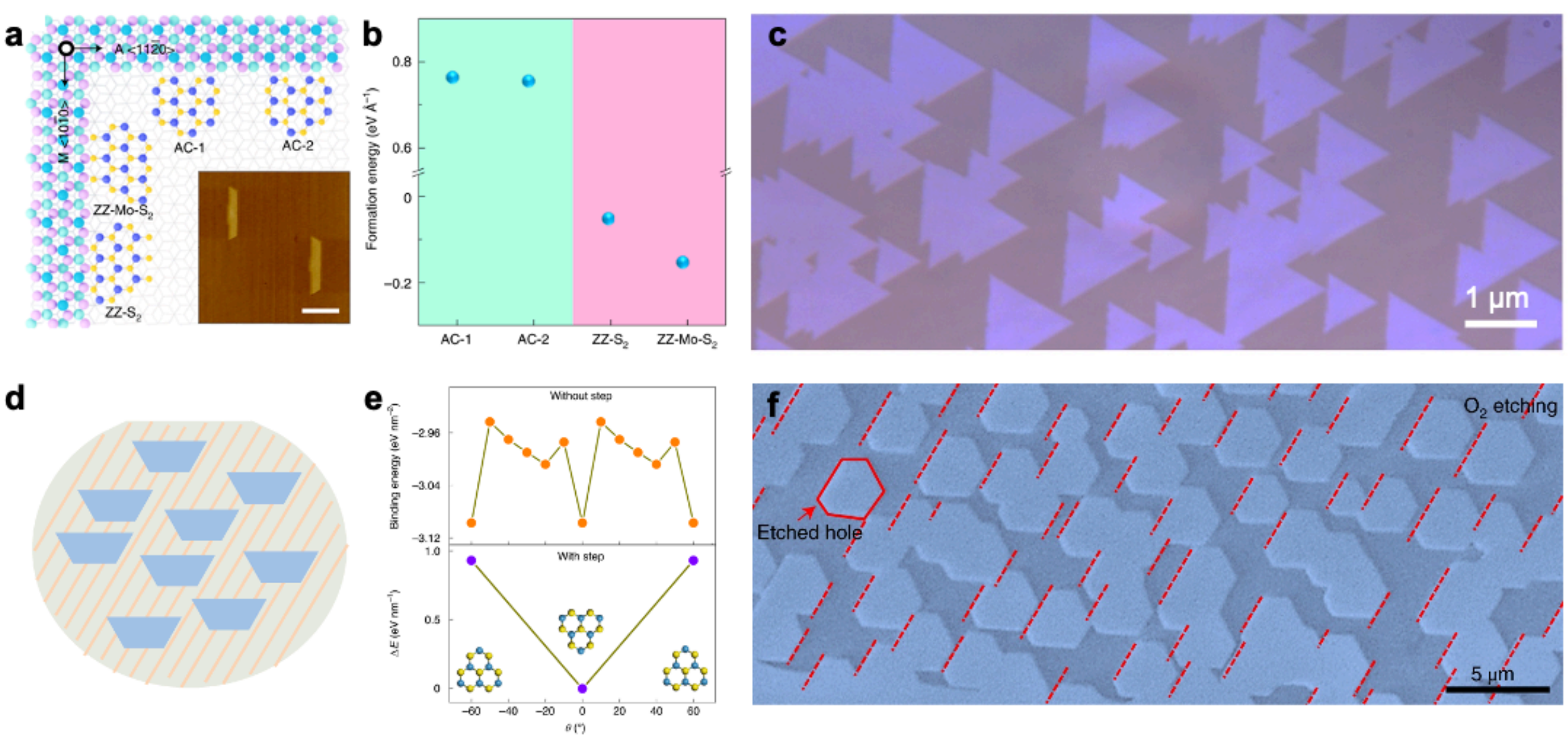

**Figure 4.** Growth strategy for single-crystal TMDs. (a) Four possible edge configurations on the C-sapphire [76]. (b) Calculated formation energy of the four edge configurations [76]. (c) Optical microscopy image of aligned $MoS_2$ domains grown on a C/A substrate. Reproduced from [76]. (d) Schematic of the growth process of $WS_2$ grains on A-sapphire [77]. (e) DFT calculations of the binding energies of a $WS_2$ triangular island on the a-plane sapphire surface without (top) and with (bottom) step edges [77]. (f) SEM image of $WS_2$ film after the $O_2$ etching. Reproduced from [77].

## 3. Growth of 2D bi- and multi-layer single crystals

In the following sub-sections, we will review the strategies developed for the growth of 2D multilayer single-crystals. We will proceed by first dealing with aligned stacking of graphene and TMDs, that is the structures in which the Nth layer is crystallographically oriented with the previous one, or to which a transformation of its point symmetry group has been applied. We will then discuss the so-called twisted configurations, and finally examine heterostructures of two material types.

### 3.1 Growth of aligned 2D bi- and multi-layers

Bernal-stacked (AB) [82] bilayer graphene (BLG) – possibly the simplest 2D multilayer one can think of – strikingly differ from its monolayer counterpart [83]. BLG features parabolic band touching points at the Brillouin zone corners (differently from the linear dispersion of graphene), leading to suppressed tunneling of normally-incident carriers [84], which can be gapped by applying a vertical electric field [85, 86]. The associated electric-field-tunable high density of states (DOS) makes BLG a fertile ground for investigating strong electronic correlation [87], which promotes magnetism and superconductivity [88] in a carbon-only bilayer. For $N \geqslant 3$, in addition to Bernal stacking, graphene multilayers can assume the so-called rhombohedral (ABC) configuration [89], in which every atom has a nearest neighbor from an adjacent layer either directly above or underneath it (in a trilayer, this can be seen as the result of a shift of the topmost layer along the armchair direction). ABC stacking is metastable, meaning that it occurs with lower probability with respect to Bernal stacking [89], and tends to convert to it under various physical mechanisms [90, 91]. Efforts to stabilize ABC multilayers [92, 93] are strongly motivated by the unique electronic properties recently discovered [94-98]. In TMDs, dramatic changes of the electronic structure arise upon increasing the number of layers. The bandgap shrinks and changes from direct to indirect with increasing the number of layers. The electronic and optical properties of these materials are not only affected by the number of layers, but also by the way they are stacked together and aligned with respect to each other.

CVD growth of graphene films [99] often results in bi- and multilayer domains [100]. Although such domains can be expanded to wafer-scale coverage [101], technological application of graphene multilayers requires high crystalline quality and control of the stacking order over device-compatible areas. Due to the small domain size and the detrimental effect of grain boundaries on electrical transport [102], polycrystalline films do not fulfil these requirements: hence the necessity of developing efficient strategies to obtain high-quality multilayer single-crystals.

Few-micron-sized AB BLG single-crystals can be obtained by a two-step approach, with epitaxial deposition on top of an existing monolayer graphene on Cu [103], resulting in >30% coverage of bilayer. To overcome the self-limiting mechanism of graphene growth on Cu substrate [99], a nearby uncovered Cu catalyst provides carbon fragments flowing downstream. Alternatively, a very high $H_2/CH_4$ ratio can be used to expose part of the upstream Cu surface of the growth foil, allowing to obtain tens-of-microns-sized single-crystals [104]. Generally speaking, the growth of large single-crystals is crucially favoured by reducing the density of nucleation centres, obtained via Cu surface treatments such as electropolishing and annealing [37]. Following this principle, it was found that pre-growth thermal annealing in Ar gas, combined with high $H_2/CH_4$ ratio and low pressure (~1 mbar), efficiently promotes the growth of large-area AB BLG [39]. These parameters optimally suppress nucleation, leading to BLG single-crystals with lateral size up to 300 μm and ~$10^4$ $cm^2$ $V^{-1}$ $s^{-1}$ carrier mobility on $SiO_2$ substrates. Comparable areas (300-550 μm) are obtained making use of a Cu pocket (that is, a folded oxygen-rich Cu foil with crimped edges) [105]. In this case, the second graphene layer grows underneath a continuous single-layer on the outer surface (as revealed by isotope-labelling experiments), taking advantage of carbon atoms diffusing from the interior of the pocket throughout the Cu foil. Pocket-grown BLG displays excellent electrical transport properties after encapsulation in hBN, including a ~100 meV band gap under a vertical electric field. Moreover, Bernal-stacked bi- and trilayers single-crystals can extend to centimetre-scale when grown by CVD on Cu/Ni alloy [106]. The growth substrate is prepared by plating Ni on both sides of a Cu(111) foil, followed by thermal annealing during several hours. It was shown that the Ni concentration in the alloy determines the solubility of carbon atoms in the substrate, which critically influences the segregation process governing the BLG growth.

Growth of ABC-stacked multilayers with stripe-like structure was early demonstrated on Cu foils [107]. Making use of curvature-engineered Cu/Ni gradient alloy, ABC trilayer can be grown with yield up to 59% of the total trilayer areas [108]. The Cu/Ni-graphene interaction crucially controls the occurrence of ABC stacking, which can be optimized by prolonging growth times (~200 min). The dimension of the ABC stripes increases (up to 1-2 μm width) after polymer-mediated transfer to $SiO_2$, due to domain walls annihilation after removal from the catalysts. Moreover, ABC is preserved after transferring on hBN and field-effect transport measurements in double-gated devices indicate the conservation of the fragile rhombohedral phase after standard device processing. ABC graphene up to 9 layers can be produced by CVD on the backside of suspended Cu foils [109], resulting in a concentric pyramidal sack of hexagonally-shaped aligned layers (figure 5(a)). Within such multilayers, the ABC stacking appear in stripes of area up to ~50 μm$^2$, alternating with Bernal-stacked parts (figure 5(b)). The morphology of the ABC stripes correlates with the corrugated surface of the Cu foil after growth, indicating that Cu step bunching contributes to the stabilization of ABC stacking. The bending of multilayer graphene over the Cu steps could activate interlayer slipping [110], which, depending on the relative orientation with respect to the graphene crystallographic directions, can either promote or suppress the rhombohedral phase, in agreement with the results of Refs. [92, 93].

The growth of thick 2D materials is particularly difficult, since they bond in the out-of-plane direction by weak vdW interaction. Recently, an isothermal dissolution–diffusion–precipitation strategy has been proposed by Zhang *et al* to grow thick graphite films [111]. They placed a solid carbon source on one surface of the single-crystal nickel (Ni) foil to enable the continuous supply of carbon atoms at high temperature. In this design, carbon atoms from the solid carbon source will isothermally dissolve into, and then diffuse through the metal foil, finally precipitate on the opposite surface of the Ni to form graphene layers continuously. With this method, they successfully obtained single-crystal graphite films with thickness up to 100,000 layers. And the obtained graphite is found to be of ultra-high quality, even better than that of natural graphite, highly oriented pyrolytic graphite (HOPG) and kish graphite.

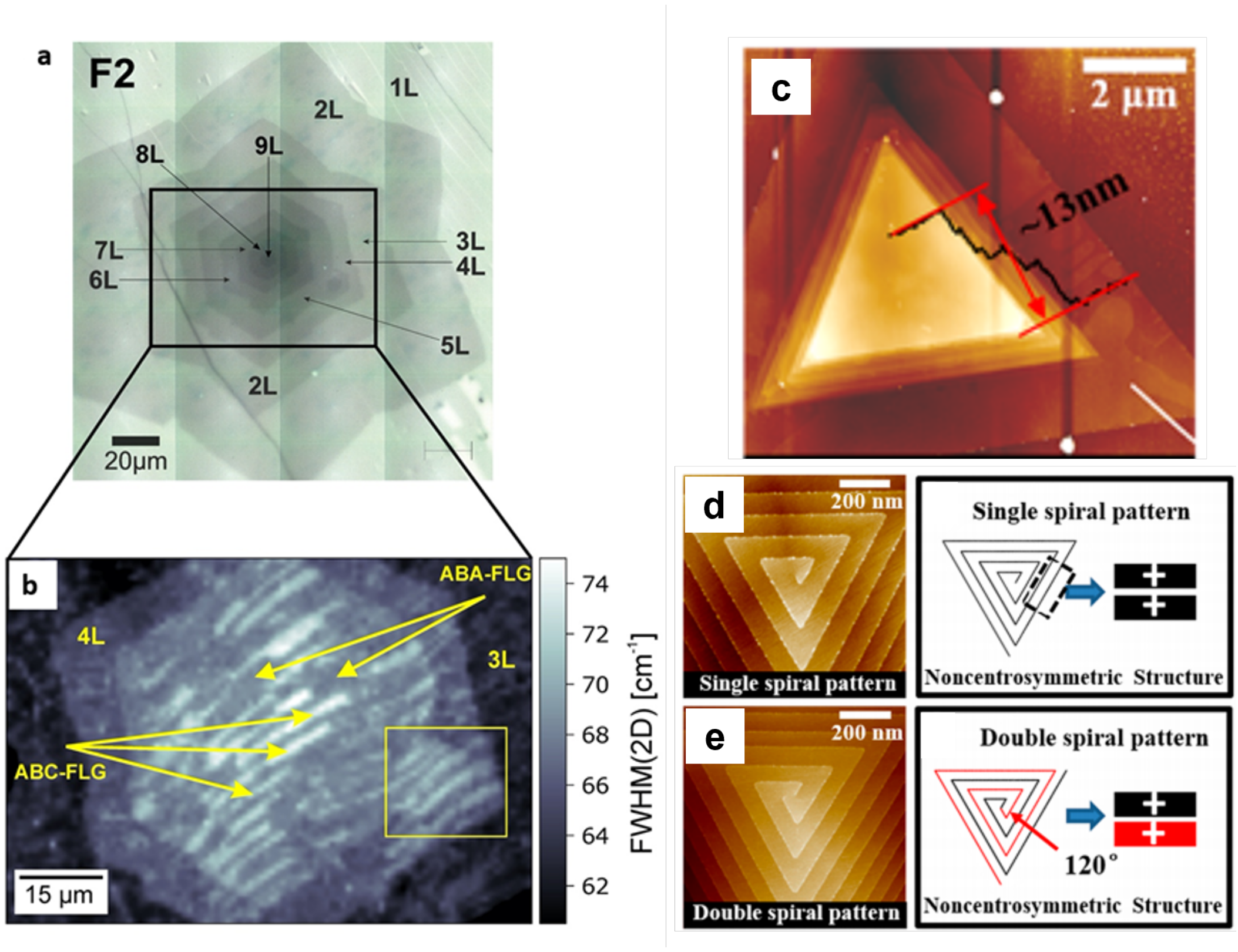

**Figure 5.** (a) Optical microscopy image of 9-layer graphene single-crystal "pyramid", after being transferred from Cu to $SiO_2$ [109]. (b) Scanning Raman microscopy image of the area highlighted in panel a, demonstrating the existence of ABC stripe-like domains, on the basis of the width of the Raman 2D peak. The ABC stripes continuously extend across thickness steps in the multilayer crystal. Reproduced from [109]. (c) AFM topography map of a wedding-cake type multilayer $WS_2$ [112]. (d, e) AFM topography of single (d) and double (e) spiral $WS_2$ multilayers. Reproduced from [112].

Chemical vapor deposition has proven to be an efficient way to successfully obtain 2D TMDs onto various substrates. The precursors for the growth of TMDs are used either in solid or in liquid form. CVD by liquid precursors can yield large crystals of the order of several hundreds of micrometres [113]. Instead, the size of the crystallites starting from solid precursors is typically of the order of the micron [114], or lower, but the quality of the crystal is excellent [114, 115]. Recent progresses in solid precursors CVD growth have allowed the demonstration of single crystals of TMDS with lateral size of hundreds of micrometres [116, 117]. Furthermore, using solid precursors implies the absence of reaction by-products at the interface with the substrate or on the surface, leaving the material atomically clean [115]. Reducing the nucleation density promotes the growth of bi- and multilayers through the Vollmer-Weber mechanism, and for hexagonal structures, the AA-alignment, where the metal and the chalcogen of the two layers are in an eclipsed configuration is the one occurring with the highest frequency [118]. The AB alignment sees a 60° rotation between the unit cells of the two layers and the AA-stack frequency is observed to be about 12% [119]. Heterobilayers of TMDs can still be grown via CVD. The growth is usually divided into a high-temperature step, where the lower layer ($WS_2$) is grown and a low-temperature step, where the top layer ($MoS_2$, $WSe_2$) is grown [120]. The CVD approach, in its various declinations, allows for the realization of heterostructures and devices up to the wafer scale [120], besides avoiding all the contaminations deriving from the use of polymer-assisted transfers. Within a scalability perspective, the approach which has gained the highest attention for its implementation in industrial pilot lines is metal-organic CVD (MOCVD). The approach might imply an economical setback. However, it yields the largest crystals so far, with wafer size coverages and controlled thickness, given the high reproducibility of the process allowed by the fine dosage tunability of the precursors [121, 122].

If the supersaturation condition is fulfilled also for $N > 2$, one can grow multilayer TMDs. The most common multilayer configuration is the so-called wedding-cake, where each layer is AA-stacked on top of the other in a pyramid-like fashion, as the structure shown in figure 5(c). However, Fan *et al* show that during the growth, a large variety of defects can occur [112]. Line defects, such as screw dislocations are very common and lead to rotated multilayer structures, as exemplified in figure 5(d) and (e). The structures shown in figure 5(d) and (e) exhibit 0°, 60° and 120° rotation angles, as they are driven by the geometry of the dislocation, i.e., they can still be considered aligned multilayers.

### 3.2 Growth of twisted 2D bi- and multi-layers

By imposing a twist angle between the stacked 2D layers, one can further tune the emerging physical properties [24]. While twisted bilayer graphene (TBG) in the large-angle limit behaves as two electronically decoupled graphene sheets with giant interlayer capacitance [123-126], small-angle TBG (below ~3° for what concerns field-effect experiments) features strong interlayer coupling and a long-range moiré superlattice [127, 128]. The spatial modulation of the interlayer hopping imposed by the moiré determines band flattening at finite magic angles [129], as recently revealed by groundbreaking experiments [130, 131]. The degree of engineering resulting from the twistronic approach [132] is unprecedented and holds great potential for fields such as quantum technologies [133]. Thus, efforts toward up-scaling of TBG are of high current interest [134]. Magic angle conditions for multilayers have been recently identified [135], in conjunction with unconventional superconducting phases [136-139]. TMDs-based moiré superlattices (resulting from either twist angle or lattice mismatch) are an ideal platform for novel physical phenomena, such as moiré excitons [140], high-temperature exciton condensation [141], correlated insulators [142], the quantum anomalous Hall effect [143]. With the huge library of combinable TMDs at disposal, the field of twisted TMDs holds comparably exciting perspectives to those of TBG [24].

CVD-grown graphene films can present abundant multilayer domains with variable interlayer twisting [144], however, due to lack of uniformity, high concentration of boundaries and uncontrolled orientation, they are typically not suitable for technological applications. Large-area twisted bilayer graphene single-crystals with uniform orientation, can be obtained by several approaches [145-151]. The fact that a dominant yield of twist angle is commonly observed at 30° (figure 6(a)) is understood based the interaction between graphene and the growth substrate at the nucleation stage [146, 147]. Owing to a docking mechanism, graphene preferentially aligns with a straight edge with atomic steps on the Cu surface. When considering the coupled graphene-Cu system, the two possible alignments at armchair and zig-zag edges are approximately isoenergetic (figure 6(b)). Therefore, despite TBG not being a stable freestanding configuration, CVD growth on Cu results in comparable amounts of 30°-TBG and AB BLG, both on Cu and Cu/Ni alloy [148]. Moreover, the same mechanism applies to larger number of layers, leading to a variety of possible interlayer configurations, including mixed Bernal-twisted ones [147, 149]. A two-step process for efficient 30°-TBG growth on Cu(111), with thermodynamically controlled nucleation and kinetically controlled growth, is reported in Ref. [149]. 30°-TBG is of high current interest as it realizes a dodecagonal quasicrystal [146, 149], a system lacking of translational symmetry, although presenting dodecagonal rotational symmetry, which results in exotic photoemission properties [152, 153]. The CVD-grown TBG quasicrystals are stable upon transfer from Cu to $SiO_2$, as well as encapsulation in hBN [146], allowing the fabrication of field-effect devices with ultra-high carrier mobility (as high as $1.9 \times 10^5$ $cm^2$ $V^{-1}$ $s^{-1}$ measured at room temperature) [126].

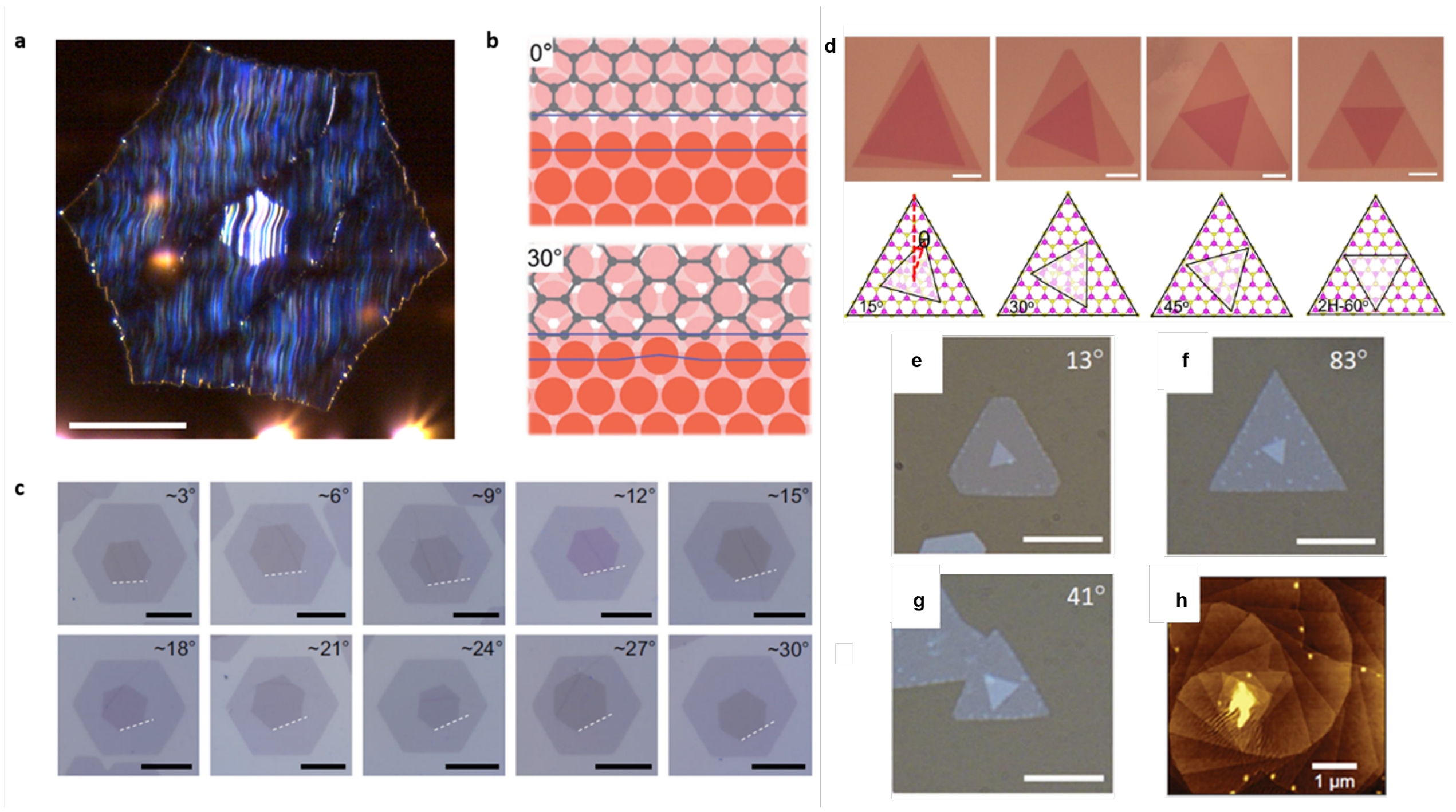

**Figure 6.** (a) Dark-field optical image of 30°-TBG grown on Cu foil. The scale bar is 50 μm. Reproduced from [146]. (b) First-principle calculations of the coupled graphene-Cu step interface, for 0° and 30° rotation (docking of a zig-zag and armchair edge, respectively). At 30°, the Cu-graphene interface remains almost as straight as for 0°, thus identifying the 30°-rotated configuration as an additional energy minimum. Reproduced from [147]. (c) TBG with a variety of twist angles obtained by hetero-site nucleation. The twist angles are identified based on the orientation of straight edges of the hexagonal crystals (white dashed lines), and confirmed by TEM experiments. Scale bars are 10 μm. Reproduced from [150]. (d) twisted $WS_2$ bilayers. Reproduced from [154]. (e-g) Optical images of $WS_2$ bilayers with different twist angles. Reproduced from [155]. (h) Supertwisted $WS_2$ spiral structures grown on $WO_x$ nanoparticles on $SiO_2$/Si. Reproduced from [156].

Small-angle TBG typically represent a small fraction of bilayer single-crystals produced in each CVD growth cycle [145]. The hetero-site nucleation strategy introduced in Ref. [150] enables a high yield (up to ~88%) of intermediate twist-angles (i.e., between 0° and 30°, see figure 6(c)). While in the cases discussed thus far the two graphene layers share the same nucleation site [157], a controlled perturbation of the gas-flow triggers nucleation of the second layer at a different site. The different local properties of the Cu surface (especially the different orientation of atomic steps) determine a variety of possible twist-angle outcomes. Very recently, Liu *et al* developed an angle replication strategy to grow centimetre-scale TBG with arbitrary twist angles. They tuned the twist angles of TBG by predesigning the rotation angle of the two single-crystal Cu(111) foils, a centimetre-scale Cu/TBG/Cu sandwich structure can be formed by precisely controlling the temperature and the TBG can be isolated by a custom-developed equipotential surface etching method [151]. Despite these promising progresses, artificial stacking mediated by hBN [158] is still to be considered the method-of-choice to access the small-angle limit of TBG, also in the case of samples based on CVD-grown graphene single-crystals [134].

The electronic and optical properties of TMDs are extremely sensitive not only to the number of layers, but also to the mutual orientation of the stacked crystals. A continuum of angles associated to the emergence of a flat band and a rich phase diagram with several quantum criticalities, ranging from quantum anomalous Hall effect to ferromagnetic order, has been predicted for moiré superlattices realized with a $WSe_2$ bilayer [159]. The twist angle between the layers can be controlled in two main ways: via transfer of one crystal onto another or via a single-step CVD process. The transfer method is still the most widely utilized and it allows for finer control of the twist angle [118]. In a single-step CVD process, the twist angle can be controlled by acting on the growth recipe, with the aim of reducing the nucleation density and favor the onset of multilayers. In addition, molecular hydrogen can be flown in the reactor for reducing the reagents and promoting the chemical reaction between them. We display in figure 6(d) the result obtained by Shao *et al* by adding Sn to the reaction, to reduce the energy of the rotated stack and favor the growth of bilayers with angles other than 0° and 60° [154]. Another approach was followed by Zheng *et al*, where they control the amount of evaporated precursor, by moving the crucible near or far from the hot zone of the furnace by means of magnets [155]. The result of such an approach is displayed in figure 6(e) and (g). In their work, Zhao and coworkers show how to realize multilayer stacks of TMDs with variable angles, with constant dephasing, exploiting the intrinsic non-euclidean geometry provided by tungsten oxide nanoparticles dispersed on a $SiO_2$/Si substrate (figure 6(h)) [156].

### 3.3 Growth of single-crystal heterostructures

2D materials allow engineering of tailored and versatile functional materials by combining them in heterostructures.

Lateral heterostructures are formed by coplanar 2D TMDs, touching each other along linear junctions [160]. These systems are predicted to exhibit highly tunable optical properties [161], as well as to be promising for applications in energy storage and quantum technologies [162]. Despite the fact that the control of the growth parameters to realize atomically sharp linear heterojunctions has still to be optimized, several notable results have been obtained for semiconductor junctions of $WS_2/MoS_2$ and for metal-semiconductor junctions between 1T' and 1H $MoTe_2$ [162, 163].

The synthesis of vertical heterostructures is a more mature process. The properties of such heterostructures are heavily affected by the degree of cleanness of the interlayer interfaces, as the interaction between the atomic orbitals would be otherwise hindered. This was first realized by placing graphene on top of an ultra-flat crystal of hBN. On an atomically flat substrate, the electronic properties of graphene can be fully unleashed and by controlling the twist angle between the two materials, the degree of interaction can be tuned, in a way that allows to access previously unexplorable physical regimes [164]. A very promising approach for the realization of vertical hetero-stacks, is to directly grow the second and successive layers, on the previous 2D material, without the need for an *ex-situ* transfer. Direct fabrication of the most common heterostructures, that of graphene and hBN, has been thoroughly investigated in the last years. Significant progresses have been made in this direction in the last decade: from demonstrating the growth of micrometer-scale single-crystals of graphene on exfoliated hBN [165] to the more recent report of scalable multilayer hBN on epitaxial graphene on SiC [166].

Graphene has an excellent chemical and thermal stability and has $C_{6V}$ symmetry. It therefore makes a perfect substrate to grow hexagonal TMDs [167, 168]. The CVD growth of TMDs directly onto graphene generally yields monolithic structures with atomically clean interface [115, 169-171] (if performed directly on graphene growth substrate), which hence improves the charge transfer efficiency and minimizes the losses due to defect trapping [172, 173]. $WS_2$ on graphene is certainly one of the most widely investigated vdW heterostructures. The monolithic structure sees $WS_2$ mostly aligned along the graphene's crystalline directions (figure 7(a)-(c)) with atomically clean interfaces [115]. The $WS_2$ crystal is locked into a commensurate configuration with the graphene lattice, but a very small perturbation is enough to set the system into an incommensurate state and let the $WS_2$ crystal slide over graphene until it reaches a defect that can halt it. This superlubric behavior might have important applications in the field of nanomechanics, as it was observed to occur not only in UHV conditions, but also in air [169]. As visible from figure 7(d), when a $WS_2$ flake is perturbed by the passage of an AFM or STM tip, the induced vibrational excitations make the flake wiggle and rotate until it escapes the commensurate state and then it starts converting its energy into translational movement.

CVD-grown vdW heterostructures exhibit quite a wide distributions of possible resulting twist angles between the 2D materials. However, the symmetry relations between the substrate and the overlayer is still key in defining the mutual orientation of the layers [168]. This implies that for graphene on hBN and hexagonal TMD on graphene or hBN the most common orientation will be 0°, 30° and 60° [165, 174], together with a uniform distribution of minority domains, randomly oriented [115].

Defects in the substrate are critical in determining both the number of layers and the quality of the grown material, as exemplified by figure 7(e) and (f). The alignment and the size difference of the lattice parameters is reflected in the reciprocal space, where the system exhibits very efficient ultrafast charge transfer from graphene into the $WS_2$ valence band [172] and it can be exploited for efficient photodetector devices [175, 176]. The variety of electronic and optical properties amongst TMDs together with the fact that graphene is prone to proximity-induced effects, make the fabrication of these hybrid structures highly appealing.

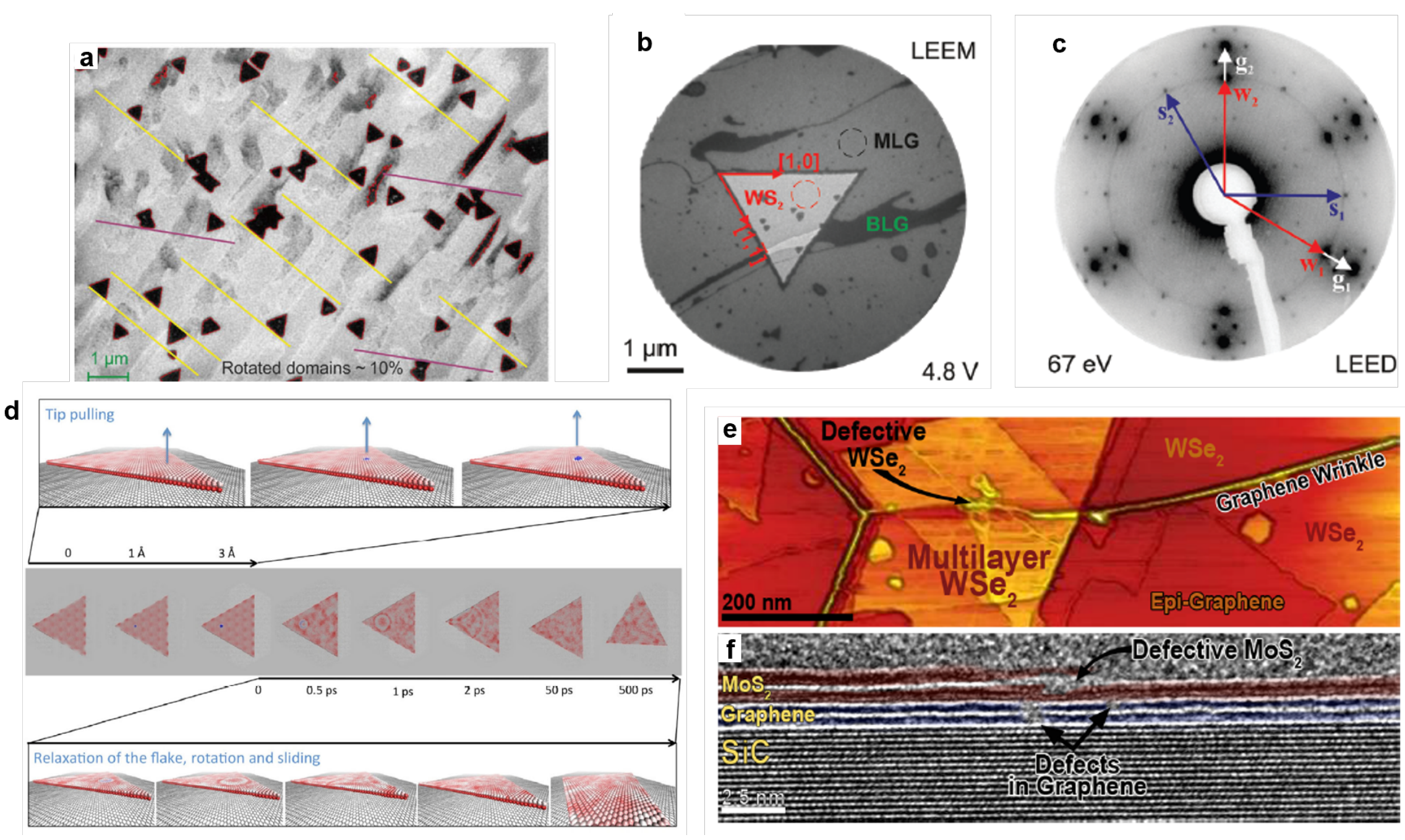

**Figure 7.** (a) SEM image of $WS_2$ crystals directly grown on graphene on 6H-SiC(0001) [169]. (b) LEEM micrograph of a $WS_2$ crystal on graphene, highlighting the crystal orientation. Reproduced from [115]. (c) LEED of $WS_2$ on epitaxial graphene, showing the reciprocal lattice vectors of graphene (white), $WS_2$ (red) and SiC (blue). Reproduced from [115]. (d) simulations of the superlubric behavior of $WS_2$ on graphene. Reproduced from [169]. (e) Scanning tunneling micrograph of $WSe_2$ grown on epitaxial graphene, demonstrating where wrinkles (or other defects) in the graphene occur; multilayer growth dominates the growth morphology [167]. (f) Transmission electron micrograph of $MoS_2$ grown on epitaxial graphene demonstrating the impact of defects in the graphene "substrate" on $MoS_2$ layer formatting. Reproduced from [167].

## 4. Applications of 2D single crystals

In this section, we will review the development of electronic and optoelectronic devices based on 2D single crystals. In principle, CVD grown monolayer sample naturally features higher defect densities than bulk crystal due to the growth environment and mechanism. Meanwhile, the inevitable transfer process limits the interface cleanliness, leading to the enhanced scattering probability. Although single-crystal growth and ultraclean transfer techniques have been developed, they are still in the early-stage with limited popularization. Therefore, the majority of prototype devices studied are still based on exfoliated samples.

### 4.1 High-performance field-effect transistors

As the essential building block of the modern integrated circuit, a high-performance FET attracts wide attention from the 2D community. This section mainly discusses the development of high-performance 2D FETs by reducing contact resistance, improving carrier mobility and on-current.

Field-effect mobility is one of the critical parameters of 2D FETs, which is greatly influenced by intrinsic and extrinsic factors. Kaasbjerg *et al* theoretically calculated the high intrinsic mobility of monolayer $MoS_2$ ( up to 410 $cm^2\ V^{-1}\ s^{-1}$ at room temperature) that is confined by the phonon scattering merely [177]. However, this ideal performance cannot be achieved due to the extrinsic effects of charge impurities, traps and defects. Firstly, charge impurities inevitably appear in the interface between semiconductors and the dielectric layer. The short interaction distance between charge impurities and conduction electrons in 2D semiconductors leads to more severe performance degradation than in bulk semiconductors. Secondly, atomically thin 2D semiconductors are unique "interface material", which are sensitive to the carrier scattering centres from the environment and fabrication process. These factors limit the early study of TMD mobility (below 10 $cm^2\ V^{-1}\ s^{-1}$) [178]. Therefore, an ultraclean surface condition is essential to reach high carrier mobility. Lee *et al* encapsulated the $MoS_2$ channel by hBN, as shown in figure 8(a) [179]. hBN is a clean dielectric without dangling bonds, which protects the $MoS_2$ surface from moisture and adsorbents in the ambient. Compared to bare 2D FETs, the mobility of hBN encapsulated FETs increased by an order of magnitude [180, 181].

Moreover, atomic and structural defects also affect the transport behaviour, resulting from the defect-induced scattering and the hopping-dominated transport mechanism [182]. Much effort has been devoted to mobility engineering since the isolation of TMDs. While researchers mainly focus on high-quality mechanically exfoliated TMDs, their small crystal size limits their large-scale integration into industrial applications. CVD method holds promise for fabricating large-scale 2D materials, but it usually suffers from a high density of grain boundaries, dislocations and vacancies, which will degrade the corresponding device performance. There is an urgent need for epitaxial wafer-scale 2D single crystals. Recently, wafer-scale monolayer $MoS_2$ single crystals has been demonstrated on sapphire substrates, and the FETs fabricated with this high-quality $MoS_2$ achieved a mobility of 102.6 $cm^2\ V^{-1}\ s^{-1}$, which is almost the highest of monolayer CVD $MoS_2$ (figure 8(b)) [76]. Figure 8(c) reveals that the mobilities of TMDs vary from 10 to over 100 $cm^2\ V^{-1}\ s^{-1}$, as engineered by numerous methods such as thermal annealing, self-assembled monolayers modification, and dielectric engineering [177, 178, 181, 183-206]. While the current research results have reached the requirement of the International Roadmap for Device and Systems (IRDS) 2034, further exploration is required to reach the intrinsic limit.

For bulk semiconductors, doping the contact region is a general approach to optimizing the resistance. However, substitutional doping usually damages the lattice and induces defects in 2D materials, thus is not a suitable method. A feasible strategy is to match the contact metal's work function with the semiconductor's electron affinity. However, metal deposition usually introduces defects, strain, disorder and metal diffusion, as shown in figure 8(d). The fabrication process of FETs dramatically influences the properties of 2D materials and always leads to Fermi-level pinning, which blocks the modulation of Schottky barrier height through metal work function [207]. vdW contacts can avoid damage to 2D semiconductors by introducing a molecular layer or a vdW gap to form a sharp interface. In figure 8(d), Liu *et al* reported a gentle metal integration approach to form an atomically flat interface between 2D semiconductors and metals, creating a Fermi-level pinning-free contact. The ideal metal-semiconductor junctions ensured high electron and hole mobility of 260 and 175 $cm^2 V^{-1} s^{-1}$, respectively [208]. However, the current density of this strategy is still limited by tunnelling transport. A semimetal-semiconductor contact could suppress metal-induced gap states (MIGS) and reduce contact resistance, further improving contact performance. As illustrated in figure 8(e), Shen *et al* have demonstrated an excellent ohmic contact formed on various monolayer TMDs. By directly evaporating bismuth (Bi) onto $MoS_2$, $MoS_2$ was degenerately doped and free of the Schottky barrier. Thus, the monolayer $MoS_2$ device shows extreme performance with a contact resistance of 123 Ω μm and an on-state current density of 1.135 mA $μm^{-1}$ [202]. Appropriately increasing the layers of 2D materials can further improve device performance without producing short-channel effects. Based on previous work [208], Liu *et al* reported that bilayer CVD $MoS_2$ FET with Bi-contact achieved the mobility of 122.6 $cm^2 V^{-1} s^{-1}$ and an on-state current of 1.2 mA $μm^{-1}$ [209]. This strategy benchmarks the performance of TMDs FETs and demonstrates the potential of 2D materials to extend Moore's Law.

According to the Landau principle, the quantum limit of $R_c$ can be obtained by

$$R_{c,min}W = \frac{h}{2q^2} e^{2k_0 d} \sqrt{\frac{\pi}{2n_s}}$$

where $W$ is the channel width, $h$ is Planck's constant, $q$ is the unit charge, $k_0$ is the decay constant, $d$ is the distance of the vdW gap between the metal and TMDs layers, and $n_s$ represents carrier concentration in the channel. Figure 8(f) established contact resistance as a function of carrier concentration with various contact modes, and the dashed line represents the quantum limit of $R_c$. Research of TMDs FETs with different contact technologies are plotted in figure 8(f) [183, 184, 191, 202, 208, 210-220]. Bi-$MoS_2$ FETs showed the lowest contact resistance with comparable performance to the bulk semiconductors, which meet the IRDS 2034 and approach the quantum limit.

Another critical parameter, a sizeable on-state current, is not only determined by the aforementioned high mobility and low contact resistance but also related to other factors. Specifically, the high electric field effect causes carriers to reach the saturation velocity ($v_{sat}$), in which the on-state current tends to be saturated. Here, $v_{sat}$ is given by

$$v_{sat} = \sqrt{\frac{8E_p}{3\pi m}}$$

where $E_p$ and m are the optical phonon energy and the carrier effective mass, respectively. High saturation current can be achieved by selecting 2D materials with low effective mass and high optical phonon energy. Furthermore, efficient gate control enables higher operating current, which requires a thinner gate dielectric with stronger electrostatic control capabilities. The loss of effective gate control capability in thin $SiO_2$ provokes the development of the high-$k$ dielectric ($HfO_2$, $Al_2O_3$), which features a lower equivalent oxide thickness. Moreover, Joule heat build-up at high currents will damage the device and limit the ultimate performance of transistors, in which the thermal dissipation of FETs also should be considered.

By optimizing the factors mentioned above, ultra-high-performance 2D FETs can be obtained, which can be used in numerous application scenarios, such as integrated circuits [27], display drivers [221], in-memory computing [222], and optoelectronics [223].

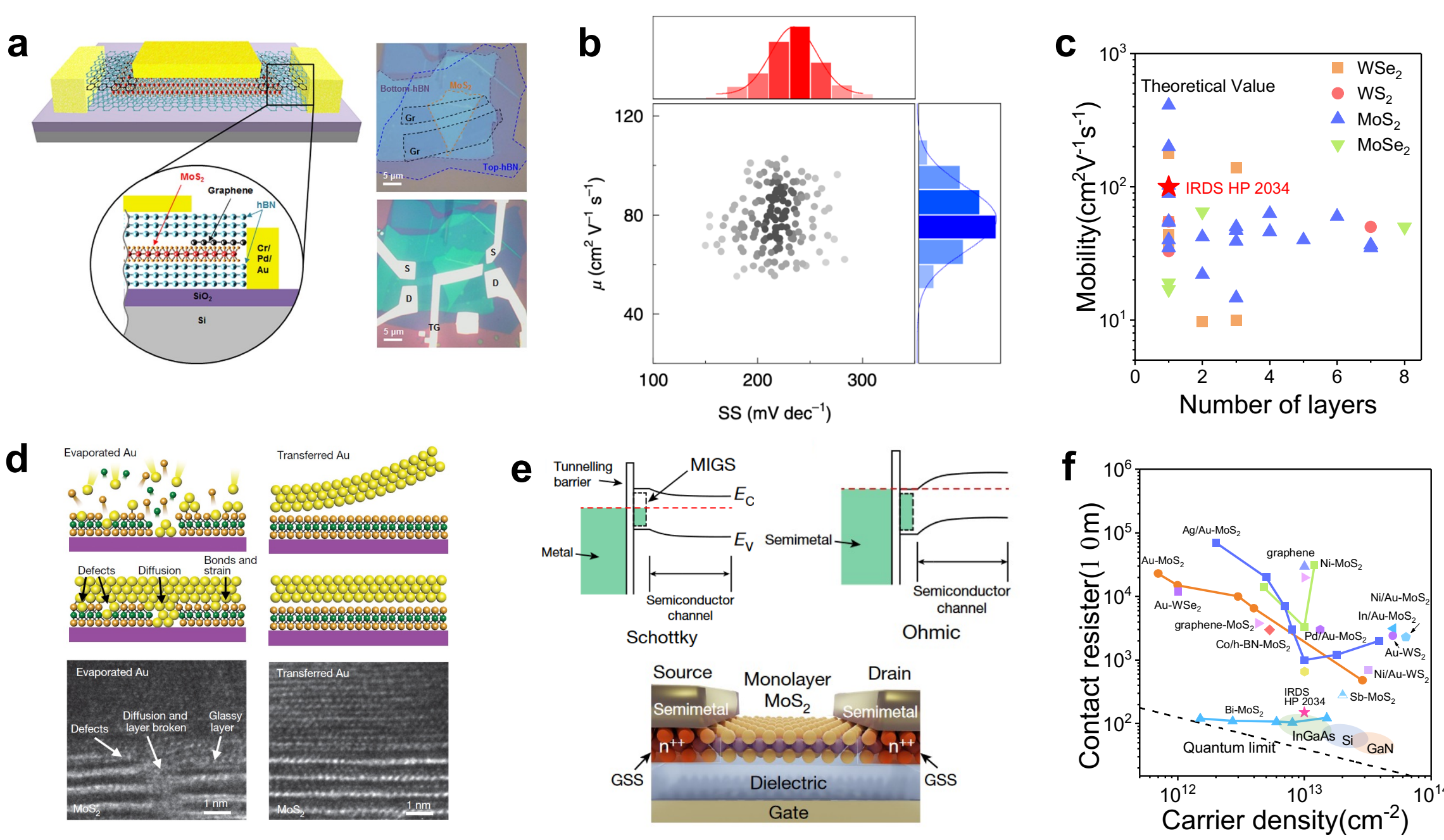

**Figure 8.** High-performance field-effect transistors. (a) $MoS_2$ field-effect transistor encapsulated by hBN. Reproduced from [179]. (b) Statistical distribution of mobility and of subthreshold swing as-grown $MoS_2$ samples. Reproduced from [76]. (c) Mobility as a function of TMDs layers [6], [11]–[38]. (d) Comparison of vdW integration junctions and deposited junctions. Reproduced from [208]. (e) The band diagram of metal-semiconductor and semimet al-semiconductor contact. Bottom is the schematic of FET with Bi-$MoS_2$ contact. Reproduced from [202]. (f) Dependence of contact resistance on carrier density in semiconductors [41], [42], [44]–[60].

### 4.2 Relevant optoelectronic devices

Optoelectronic devices that convert optical signals to electrical signals or vice versa are essential building blocks in the modern era of information and big data. So far, numerous optoelectronic devices made of 2D semiconductors have been demonstrated, showing rich device physics and application scenarios. A photodetector is the most widely studied optoelectronic device, whose most straightforward structure is the same as a FET (figure 9(a)). However, the photodetector performance varies considerably in such a simple device structure in the literature. Taking $MoS_2$ as an example, the reported responsivity and response time vary from ~$10^{-5}$ to $10^4$ A $W^{-1}$ and from picoseconds to seconds, respectively, indicating the multiple mechanisms in the photoelectric conversion process [224-229]. Figure 9(b) shows a typical transfer curve in the dark and under light, and the corresponding photocurrent, respectively [230]. The monotonous increase of $\Delta V_T$ (the negative shift of the threshold voltage under illumination) with increasing the illumination power (inset of figure 9(b)) implies the dominant photocurrent source of the photovoltaic effect, also known as the photogating effect, which contributes to a high responsivity of ~$10^3$ A $W^{-1}$. When changing the atmosphere from $N_2$ to air, $\Delta V_T$ dramatically increases from 22 to 150 V, suggesting the dominant charge trapping centre is the water/oxygen molecules between the $MoS_2$ flake and the $SiO_2$ substrate. By chopping the incident light up to 3 kHz, a much smaller but faster photocurrent process can be distinguished, which results from the photoconductive effect, i.e., the increase in conductivity due to the photoinduced excess carriers (figure 9(c)). The photoconductive gain can effectively prolong the lifetime of the excess carriers. Specifically, the holes are easily trapped in the valance band tail, leading to a long practical carrier lifetime, i.e., the photogenerated electrons can circulate several rounds between source and drain electrodes before recombination (figure 9(d)). This effect vastly increases the photoconductive responsivity from the intrinsic 0.06 A $W^{-1}$ to 6 A $W^{-1}$ [230]. Using a two-pulse photovoltage correlation technique, Wang *et al* extracted two response timescales of 3-5 and 80-110 ps, corresponding to the carriers captured by the fast and slow defects, respectively [231]. Due to the charge trapping centre at the interfaces or crystal defects, a trade-off exists between the responsivity and operation speed. Developing the appropriate interface, defect, and dielectric engineering methods (such as defect healing, degenerate doping, and screening of charge scattering centres) is essential to push the device's performance forward to the limit.

In bulk semiconductors, heterojunctions such as quantum wells possess charming properties in fundamental physics and device applications. 2D crystals are a better platform for studying the heterojunction interaction due to their immunity against the lattice-constant mismatch, enabling a broad range of material combinations. Figure 9(e) shows a schematic bilayer heterostructure for photodetection, composed of n- and p-type semiconductors, as demonstrated in numerous groups [232-237]. For example, Cheng *et al* fabricated a CVD-grown $WSe_2/MoS_2$ heterojunction, showing an excellent current rectification behaviour with an ideality factor of 1.2 (figure 9(f)). The built-in field in this heterojunction could effectively separate the photogenerated electron-hole pairs, as evidenced by the open-circuit voltage and short-circuit current under illumination. The maximum external quantum efficiency reaches 12%, with a fast response of <100 μs [238]. Moreover, InSe avalanche [239] and InSe/BP ballistic avalanche [240] photodetectors have been demonstrated, indicating the great potential of 2D single crystal or heterojunction for weak signal detection towards the single-photon limit. Besides photodetection, light-emitting devices have also been achieved in various structures, including pn junctions [241, 242], quantum well structures [243], and metal-insulator-semiconductor heterojunctions [244, 245].

Another critical group of devices is optoelectronic memories. Usually, the stored information is represented by different resistance levels, requiring the memory device to keep unchanged for a long time despite the environment, and easy to store or release charges under the electrical or optical stimulus. Lee *et al* intentionally introduced the charge traps in the Si substrate by an oxygen plasma treatment. These localized states serve as the charge trapping centre that can be electrically filled or optically released, leading to a maximum current on-off ratio of ~4700 and a long storage lifetime of over $10^4$ s [246]. Xiang *et al* further improved the optoelectronic memory performance using p-type $WSe_2$ as the channel material and defective hBN as the charge trapping dielectric [247] (figure 9(g)). Under laser illumination and negative gate pulse (-20 V), the midgap states in hBN are excited, and the photogenerated electrons flow into $WSe_2$ under the electric field (figure 9(h) program state). After switching the gate voltage back to +50 V, the electrons and holes reside in $WSe_2$ and hBN, respectively, presenting the high conductance at the reading state. The device achieves the erase state under illumination at +50 V, where the valence electrons are excited to recombine with the holes trapped in hBN, and the holes flow into $WSe_2$ to compensate for the free electrons under the electric field (figure 9(h), erase state). This device reaches up to $1.1\times10^6$ switching ratio with 130 storage states (7 bits storage), retention time over $4.5\times10^4$ s, and is colour distinguishable because the density of hBN midgap states is sensitive to the program wavelength.

Finally, a field where 2D materials might significantly contribute is that of integrated photonics for future datacom and telecom applications [248]. Indeed, to date photonic building blocks such as modulators and photodetectors integrating CVD grown single-crystal graphene have been demonstrated [61, 249]. They do not only feature high data-rates, small footprint and low power consumption, but can be also straightforwardly integrated on existing waveguide technology [250].

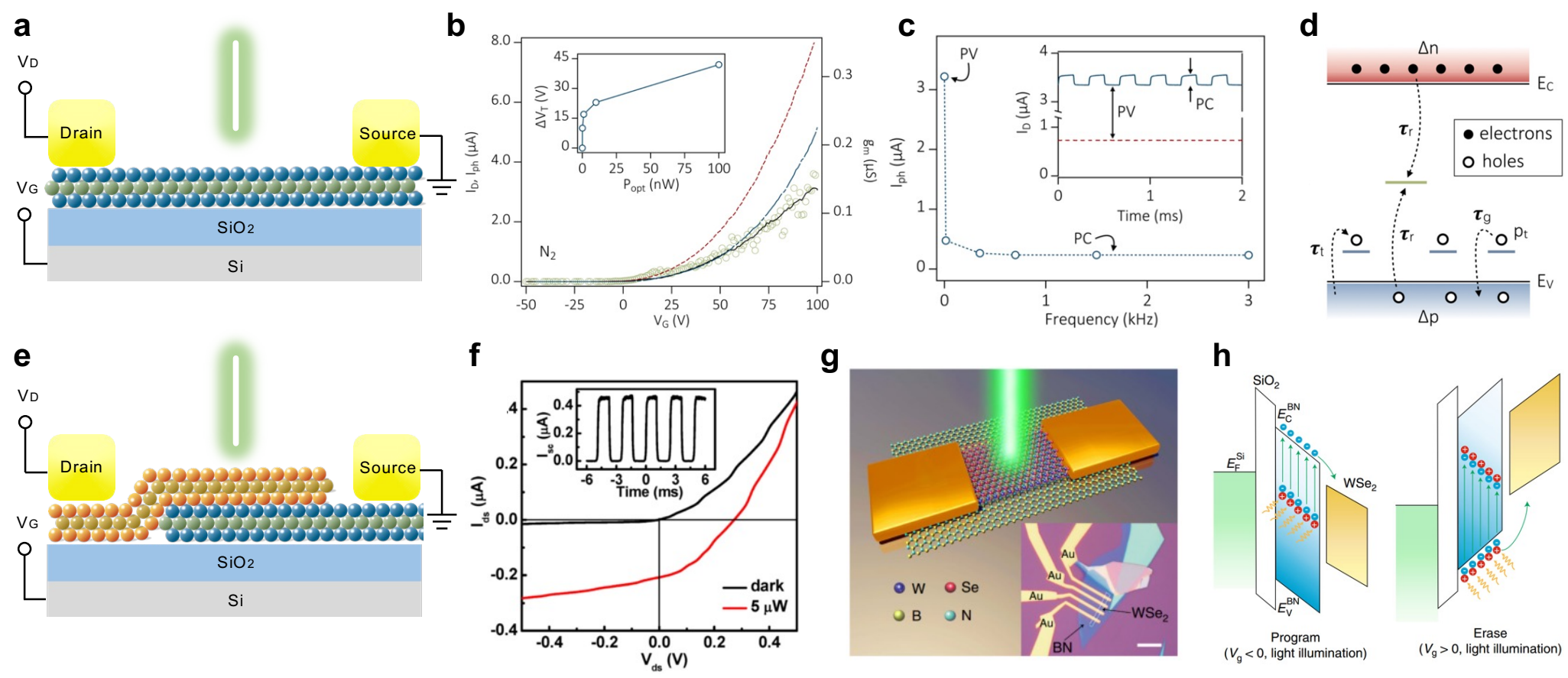

**Figure 9.** Relevant optoelectronic devices. (a) Device structure of a FET type photodetector. (b) Transfer curve performed in the dark (dash-dotted blue) and under illumination with 5 nW (dashed red). The photocurrent is shown as the solid black line. The transconductance is plotted in the same graph on the right axis (symbols). (Inset) Dependence of $\Delta V_T$ on incident optical power [230]. (c) Photocurrent versus modulation frequency. The photoconductive effect shows no modulation frequency dependence in the measured range (up to 3 kHz). (Inset) $I_d$-time traces taken in the dark (dashed red line) and under illumination (solid blue line) [230]. (d) Simplified energy band diagram that shows the main features of the charge trapping model. Reproduced from [230]. (e) The device structure of a heterojunction. (f) $I_d$-$V_d$ curve in the dark and under illumination. Inset: $I_d$-time traces of the device. Reproduced from [238]. (g) Schematic illustration of the optoelectronic memory device composed of $WSe_2$ on hBN. Inset: the optical image of the memory device. The scale bar is 10 μm [247]. (h) Working principles of the $WSe_2$/hBN optoelectronic memory under the programming and erasing processes. Reproduced from [247].

## 5. Outlook

2D materials have witnessed a decade-long research boom since their first isolation via micromechanical exfoliation. The growth techniques for 2D single crystals and their appealing applications are two key research directions in this field. Recent breakthrough progresses both on the growth and applications of 2D single crystals have been summarized in this review. However, numerous challenges need to be addressed to reach fully controllable growth and high-performance device fabrication in the near future.

The route towards market-ready fabrication of 2D-materials-based devices for the post-Moore era must foresee the growth of single crystal 2D materials over wafer-size substrates in a reliable and reproducible fashion [251]. Despite the progresses made in that direction by MBE, especially useful for magnetic and air sensitive TMDs [252], and by the growth of 2D metal-organic frameworks [253], CVD and MOCVD still maintain a dominant role for the synthesis of high-quality large-area two-dimensional crystals as they provide simplicity, cost-effectiveness, and scalability [254, 255].

Devices based on 2D materials are considered the most promising candidate for future electronic and optoelectronic technology. The wafer-level growth of single-crystal monolayer 2D materials allowed for the development of large-area integrated circuits. Doping 2D materials with specific elements could greatly broaden their application scope with huge electrical tunability. However, due to the thermodynamic phase segregation nature, a uniformly doped 2D material with controllable dopant distributions and desired doping levels is still of great challenge to be obtained.

Besides, multilayer 2D materials with higher mobility and electrostatic control are required to improve device performance further. However, the controlled growth of large-area single-crystal multilayers remains a challenging project, where the thickness control, stacking sequence, twist angle and heterostructure construction arise as the most interesting research areas. Despite the progress that have been made, the direct growth of 2D materials with controllable thickness, desired stacking order and twist angle, and designed heterostructures still deserves continuous exploration. Based on the recent progress in the field, there are several potential strategies that could possibly solve these issues, i.e., by designing new precursor supply way, introducing extra substrates or providing desired temperature gradients into the system.

In addition to the channel material itself, the transistor performance is also related to the device fabrication process, especially the work function of contact metal, the deposition of the dielectric layer, and the cleanliness of the vdW interface. At present, low-temperature deposition of metal Bi can form semimetal-semiconductor contacts with $MoS_2$, which provides an effective path to the Ohmic contact of n-type 2D FETs. However, the research on p-type Ohmic contact is still in the early stage, which significantly limits the realization of p-type 2D FETs and the development of 2D CMOS circuits. The deposition of ultra-thin high-$k$ dielectric has a great impact on device performance. High-quality dielectric layers will bring low leakage current, high device stability and low density of interface states. Although many techniques have been developed, it requires much more effort to realize sub-1nm EOT with a low interface state density comparable to the Si MOSFET. With a rigorous requirement on the interface quality, non-destructive transfer methods, selection of encapsulation layers, and interface modification of substrates require further exploration.

For 2D optoelectronics, a crucial target is pushing the device performance toward the limit, which requires a comprehensive understanding of the characteristics of interfaces and defects. The trade-off in optoelectronics, e.g., responsivity and speed for a photodetector, retention and endurance for an optoelectronic memory, must be carefully treated in evaluating the device performance. Moreover, researchers should utilize the ease of 2D heterostructure fabrication to explore new conceptual devices that cannot be realized with conventional bulk semiconductors. For example, the robust exciton with high binding energy permits the exploration of room-temperature exciton transistor and exciton-polariton lasers. These novel optoelectronics with new device physics and carrier dynamics take advantage of the unique properties of 2D materials and fill the blanks of current optoelectronics. Meanwhile, the research on 2D optoelectronics should not be limited to a single prototype device demonstration but focus on integrated devices with more advanced functions, which require a significant advance in controllable doping, wafer-scale single crystal growth, ultraclean sample transfer, and back-end-of-line integration.

To this end, wafer-scale single crystal growth, prototype device study, and the corresponding integration techniques should be studied systematically, which demands multi-disciplinary cooperation among material science, physics, electrical engineering and computer science. The academic community and the industrial sector shall cooperate closely to push the 2D materials devices toward shaping the future information industry.

**Acknowledgments**: This work was supported by Guangdong Major Project of Basic and Applied Basic Research (2021B0301030002), the National Natural Science Foundation of China (52025023, 51991342, 52021006, 61927808, 61734003, 61861166001, 51861145202, T2188101 and T2221003), Leading-edge Technology Program of Jiangsu Natural Science Foundation (BK20202005), the Key R&D Program of Guangdong Province (2019B010931001), the National Key R&D Program of China (2021YFB3200303), the Strategic Priority Research Program of Chinese Academy of Sciences (XDB33000000).